\documentclass[12pt,a4paper]{article}

\RequirePackage[l2tabu, orthodox]{nag}

\usepackage{jheppub}
\usepackage{amsthm,amssymb,amsmath,epic,float}
\usepackage{rotating,epsfig,indentfirst,array,varioref}
\usepackage{appendix,tikz,mathtools}
\usepackage{setspace}
\usepackage{tikz}
\usetikzlibrary{decorations.pathreplacing}
\usetikzlibrary{calc}
\usepackage{enumitem}
\usepackage{hyphenat}
\usetikzlibrary{positioning}
\usepackage{graphicx}  

\usepackage{jheppub}
\makeatletter
\def\@fpheader{\vspace{-.1cm}}
\makeatother

\title{\boldmath On the equivalence of Batyrev and \\  BHK  Mirror symmetry constructions }

\author[a,c,d]{Alexander Belavin,}
\author[a,b,c,d]{Boris Eremin}

\affiliation[a]{Landau Institute for Theoretical Physics, 142432 Chernogolovka, Russia}
\affiliation[b]{Skolkovo Institute of Science and Technology, 143026 Moscow, Russia}
\affiliation[c]{Moscow Institute of Physics and Technology, 141700 Dolgoprudny, Russia}
\affiliation[d]{Kharkevich Institute for Information Transmission Problems, 127994
Moscow, Russia}

\emailAdd{Belavin@itp.ac.ru,  Eremin.ba@phystech.edu}

\abstract{We consider the connection between two constructions of the mirror partner for the Calabi-Yau orbifold. This orbifold  is defined as a quotient by some suitable subgroup $G$ of the phase symmetries of the hypersurface $ X_M $ in the weighted projective space, cut out by a quasi-homogeneous polynomial $W_M$.
The first, Berglund-Hübsch-Krawitz (BHK) construction, uses another weighted projective space and the quotient of a new hypersurface $X_{M^T}$ inside it 
by some dual group $G^T$.
In the second, Batyrev construction, the mirror partner is constructed  as a hypersurface in the toric variety defined by the reflexive polytope dual to the polytope associated with the original Calabi-Yau orbifold.
We give a simple evidence of the equivalence of these two constructions.}


\begin{document} 
\maketitle
\flushbottom

\section{Introduction}
\label{sec:intro}

Mirror symmetry provides a geometrical connection between a pair of algebraic manifolds and has been studied for decades. According to mirror conjecture for a pair of Calabi-Yau threefolds $(X,Y)$ there is an isomorphism of the cohomologies
\begin{equation}
  H^{p,q}(X,\mathbb{C})=H^{3-p,q}(Y,\mathbb{C}).
\end{equation}
Calabi-Yau manifolds $X$ and $Y$, being complex and Kähler manifolds, allow deformations of complex and Kähler structures. Thus, moduli space of complex $M_C(X)$ and Kähler $M_K(X)$ structure deformations arise \cite{Candelas:1990pi}. The mirror symmetry can be interpreted as matching the  Special geometries \cite{Strominger:1990pd}, on the moduli spaces
\begin{equation}
    M_C(X) \simeq M_K(Y), \ \ \ M_K(X) \simeq M_C(Y),
\end{equation}
and the equality of their dimensions (or the Hodge numbers)
\begin{equation}
    h_{21}(X)=h_{11}(Y), \ \ \ h_{21}(Y)=h_{11}(X).
\end{equation}

In this paper we consider Calabi-Yau manifolds cut out by  quasi-homogeneous polynomial
\begin{equation}
\label{eq:CalabiYau}    
W_M(x)=\sum_{i=1}^5 \prod_{j=1}^5 x_j^{M_{ij}}
\end{equation}
in a weighted projective space
\begin{equation}
\label{eq:weighted}
\mathbb{P}^4_{(k_1,k_2,k_3,k_4,k_5)}=\Big\{ (x_1,\dots,x_5)\in \mathbb{C}^5 \setminus \{0\} \ \Big| \ x_i \sim \lambda^{k_i} x_i \Big\}.
\end{equation}
Such polynomials are also called \textit{invertible}, meaning the number of monomials is equal to the number of variables, also  the polynomial satisfy the following conditions:
\begin{itemize}
\item[(i)] the matrix $M$ is integer, and invertible;

\item[(ii)] the polynomial $W_M$ is quasi-homogeneous, i.e., there exist positive integers $k_i$ (weights of the 
projective space), and $d$ (degree), such that
\begin{equation}
\label{eq:CYcondition}
    \sum_{j=1}^5 M_{ij}k_j=d=\sum_{i=1}^5k_i, \ \ \ \forall i;
\end{equation}

\item[(iii)] the polynomial $W_M$ is a non-degenerate potential away from the origin. It follows that the $W_M$ must be a sum of invertable potentials of one of three atomic types:  Fermat, loop, or chain \cite{Kreuzer:1992bi}.
\end{itemize}
These facts provide that the $W_M$ defines a Calabi-Yau manifold. Note some useful properties of the matrix $M$, and their inverse matrix  $B=M^{-1}$. There also exist positive integer numbers $\bar{k}_j$, and $\bar{d}$ such that 
\begin{equation}
\label{eq:CYconditionT}
     \sum_{i=1}^5\bar{k}_i M_{ij}=\bar{d}=\sum_{l=1}^5 \bar{k}_l, \ \ \ \forall j.
\end{equation}
Using the quasi-homogeneity condition (\ref{eq:CYcondition}),(\ref{eq:CYconditionT}) and the definition $\sum_l B_{il}M_{lj}=\delta_{ij}$ one can obtain the relations 
\begin{equation}
\label{eq:matrixB}
    \sum_{j=1}^5 B_{ij}=\frac{k_i}{d},
\end{equation}
and
\begin{equation}
\label{eq:matrixBT}
    \sum_{i=1}^5B_{ij}=\frac{\bar{k}_j}{\bar{d}}.
\end{equation}

Calabi--Yau $X_M$ allows the deformation of complex structure. In terms of the polynomials, it is realized as a deformation of the $W_M$. Thus the full family of $X_M$ is given by zero locus of 
\begin{equation}
\label{eq:fullfamillyofX}
    W(x,\varphi)=\sum_{i=1}^5 \prod_{j=1}^5 x_j^{M_{ij}}+\sum_{s=1}^{h}\varphi_s e_s(x),
\end{equation}
where $e_s$ are also quasi-homogeneous and form the basis in the space of deformations of complex structure. The $\varphi_s$ are moduli of complex structure, and $h$ is the Hodge number of the family $X_M$.

In this paper we present a simple verification of the equivalence between
Berglund – Hübsch – Krawitz and Batyrev constructions for mirror symmetry.
Berglund and Hübsch proposed \cite{Berglund:1991pp} that the mirror partner for the hypersurface $X_M$ is related to the hypersurface $X_{M^T}$ cut out by 
\begin{equation}
    W_{M^T}(z)=\sum_{i=1}^5\prod_{j=1}^5 x_j^{(M^T)_{ij}}
\end{equation}
in another weighted projective space $\mathbb{P}^4_{(\bar{k}_1,\bar{k}_2,\bar{k}_3,\bar{k}_4,\bar{k}_5)}$. They suggested that the mirror of $X_M$ is realized as a quotient of $X_{M^T}$ by some subgroup of the phase symmetries of the $W_{M^T}$.
This approach has been generalized by Krawitz \cite{krawitz2009fjrw}. The construction starts with the polynomial $W_M$ and the hypersurface $X_M$. Let $SL(M)$ be the group 
of phase symmetries preserving $H^{3,0}(X_M)$, and  $J_M \subseteq SL(M)$ -- its subgroup   that consists of the phase symmetries induced by $\mathbb{C}^*$ action on 
$\mathbb{P}^4_{(k_1,k_2,k_3,k_4,k_5)}$.
Choose a group $G_0$ to be some subgroup of $SL(M)$ containing $J_M$. It was shown that the quotient space $Z(M,G):=X_M/G$ is a Calabi-Yau orbifold, 
where $G=G_0/J_M$.  
A similar procedure can be applied to the transposed polynomial $W_{M^T}$ to obtain Calabi-Yau orbifold $Z(M^T,G^T):=X_{M^T}/G^T$. Krawitz has shown how to choose the dual group $G^T$ such that $Z(M,G)$ and $Z(M^T,G^T)$ form a mirror pair on the level of cohomologies \cite{krawitz2009fjrw}, see also \cite{Shoemaker:2014cda,Kelly:2013lmu, Clader:2014kfa}. We describe this construction in more detail in the next section.

On the other hand, following Batyrev \cite{Batyrev:1994hm}, we can build a mirror of the original orbifold as follows.
The superpotential $ W_M $ after adding to it admissible, that is invariant with respect to the group $ G $, quasi-homogeneous monomials, that correspond to deformation of the conformal structure, defines a set of  vectors 
$\Vec{V}_a\in\mathbb{Z}^5$. 
These vectors  are  the exponents of the monomials in  $W_M$ and its admissible deformations. 
The $\Vec{V}_a$, after subtracting  the vector $\Vec{v_0}$ with components $(1,1,1,1,1)$ from each of them, begin to belong  to the four-dimensional sublattice, determined by its orthogonality to the vector $\Vec{k}:=(k_1,k_2,k_3,k_4,k_5)$.
Moreover, more precisely, the vectors $\Vec{V'}_a :=\Vec{V}_a - \Vec{v_0}$ lie in a sublattice of the $L$,  since they satisfy  additional constraints, namely, they correspond only to the $G$--invariant monomials. The lattice  points  $\Vec{V'}_a$  consist the Batyrev reflexive polyhedron $\Delta(M,G)$ for the orbifold $Z(M,G)$. On the other hand, the vectors $\Vec{V'}_a$ correspond to the edges of the $\textit{fan}$ for the dual reflexive polyhedron. 
Having the data of the  $\textit{fan}$, we build the mirror for the $Z(M,G)$, let us call it $Y$, in two steps. First,  we construct   the toric variety $T$ \cite{Batyrev:1994hm, Hori:2003ic}. On the second step, we find the homogeneous polynomial $W^Y$ whose critical locus cut out a hypersurface which is nothing but the mirror manifold $Y$.  
After that, by reducing the toric variety $ T $ to a weighted projective space $\mathbb{P}^4_{(\bar{k}_1,\bar{k}_2,\bar{k}_3,\bar{k}_4,\bar{k}_5)}$ which appears in the BHK construction, we demonstrate that the results of the two constructions coincide.
The "strong parallel between the Berglund-Hübsch-Krawitz and Batyrev  mirror symmetry", 
based on Landau-Ginzburg/Calabi-Yau correspondence,  was stated by
Clader and Ruan in \cite{Clader:2014kfa}. In this paper we  confirm exactly the equivalence of 
these two constructions by the explicit computation for the CY orbifolds connected with the invertible singularities.
  
In Section \ref{sec:2}, we fix notations and give a short review of the BHK mirror construction. 
In Section \ref{sec:3}, we formulate a version of the Batyrev approach for constructing mirror orbifolds and show its equivalence to the BHK construction.

\section{Berglund–Hübsch–Krawitz mirror construction}
\label{sec:2}

Let Calabi-Yau hypersurface $X_M$ be defined in $\mathbb{P}^4_{(k_1,k_2,k_3,k_4,k_5)}$ by zero locus of 
\begin{equation}
W_M(x)=\sum_{i=1}^5 \prod_{j=1}^5 x_j^{M_{ij}}.
\end{equation}
Taking into account the quasi-homogeneity of $W_M$  we obtain that it is invariant under the action of the group $J_M$ generated by the following action
\begin{equation}
\label{eq:QMaction}
    x_i\mapsto \omega^{k_i} x_i, \ \ \ \omega^d=1.
\end{equation}

Moreover, the polynomial $W_M$ has a larger group of diagonal automorphisms
\begin{equation}
  \text{Aut}(M):= \{ (\lambda_1,\dots,\lambda_5) \in (\mathbb{C}^*)^{5} \  |\  W_M(\lambda_1x_1,\dots,\lambda_5 x_5) =W_M(x_1,\dots,x_5), \ \forall x_i \},
\end{equation}
of order $|\text{Aut}(M)|=d$ and $J_M \subseteq \text{Aut}(M)$. The group $\text{Aut}(M)$ is generated by $q_i(M)$ which act on coordinates as
\begin{equation}
    q_i(M): x_j \mapsto e^{2\pi i B_{ji}} x_j,
\end{equation}
where the matrix $B=M^{-1}$.
Indeed, the generators of $\operatorname{Aut}(M)$ act on each term in 
(\ref{eq:CalabiYau}) as
\begin{equation}
    q_l(M)\cdot \prod_{j=1}^5 x_j^{M_{ij}} = e^{2\pi i B_{ij}M_{jl}}\prod_{l=1}^5x_l^{M_{il}}=e^{2\pi i \delta_{il}}\prod_{l=1}^5x_l^{M_{il}}=\prod_{l=1}^5x_l^{M_{il}}.
\end{equation}
In these terms, the generator of the group $J_M$ is
\begin{equation}
    \prod_{i=1}^5q_i(M).
\end{equation} 

Calabi-Yau threefold $X_M$ admits the existence of holomorphic, nowhere vanishing 3-form $\Omega$. Subgroups preserving this form $\Omega$, or, equivalently, preserving the product $\prod_{i}x_i$
are called allowable. Let $S L\left(M\right)$ be the maximal allowable group with generators $p_s(M)$,
\begin{equation}
\label{eq:maxallow}
S L\left(M\right):=\left\{p_s(M) \in \operatorname{Aut}\left(M\right) \mid p_s(M)\cdot \prod_{j=1}^5x_j=\prod_{j=1}^5x_j\right\}.
\end{equation}
The obvious fact is that $J_M\subseteq SL(M)$. 
 Consider an allowable subgroup  $G_0$ such 
that $J_M \subseteq G_0 \subseteq SL(M)$. Define the quotient group
\begin{equation}
    G:=G_0/J_M.
\end{equation}
Then we obtain Calabi-Yau orbifold $X$ as
\begin{equation}
    X:=Z(M,G)=X_M/G.
\end{equation}

For the first time, such orbifolds and their mirrors were considered for the case of 
Calabi-Yau threefold  in \cite{Greene:1990ud}.
In the general case, the Berglund–Hübsch–Krawitz construction starts from the quasi-homogeneous, invertible polynomial $W_M$ and the group $G_0$ 
\begin{equation}
\label{eq:embedding}
    J_M \subseteq G_0 \subseteq SL(M) \subseteq \operatorname{Aut}(M).
\end{equation}
The full family of the orbifold $X$ cut out by the equation $\{W^X=0 \} \subseteq \mathbb{P}^4_{(k_1,k_2,k_3,k_4,k_5)}/G$. Here the polynomial is
\begin{equation}
\label{eq:Familly}
    W^X(x,\varphi)=\sum_{i=1}^5 \prod_{j=1}^5 x_j^{M_{ij}}+\sum_{l=1}^{h_X}\varphi_l \prod_{j=1}^5 x_j^{S_{lj}},
\end{equation}
where $\varphi_l$ are moduli of complex structure deformations, and $h_X:=h_{21}(X)$ is Hodge number. The  monimials
\begin{equation}
    \label{eq:Monomials} 
    e_l:=\prod_{j=1}^5 x_j^{S_{lj}}
\end{equation}
in (\ref{eq:Familly}) are quasi-homogeneous since $\sum_{j}S_{lj}k_j=d$ and invariant under the group $G$. They belong to $G$-invariant subring of the Milnor ring  $\mathbb{C}[x_1,\dots,x_5]/ \langle  \frac{\partial W_M}{\partial x_j} \rangle$ \cite{Lerche:1989uy}. We denote the monomial with $S_{h_X,i}=1$, which  plays a distinguished role, $e_{h_X}$,
\begin{equation}
\label{eq:eh}
    e_{h_X}=\prod_{i=1}^5x_i.
\end{equation}
  Actually, the monomials $e_l$ in (\ref{eq:Familly}) are the subset in the basis of deformation of complex structure $e_s, \ s=1,\dots,h$ (see (\ref{eq:fullfamillyofX})) of the original CY family $X_M$. 
	If we denote by $\rho_s$ the generator of the group $G_0$, then we will see that  
\begin{equation}
    \rho_s \cdot \prod_{j=1}^5 x_j^{S_{lj}} = \prod_{j=1}^5 x_j^{S_{lj}} , \ \ \ l=1,\cdots,h_X.
\end{equation} 
We can define a similar (\ref{eq:embedding}) set of groups for the transposed matrix: $ J_{M^T} \subseteq G^T_0 \subseteq SL(M^T) \subseteq \operatorname{Aut}(M^T)$.
 Recall, that the polynomial
\begin{equation}
    W_{M^T}(z)=\sum_{i=1}^5 \prod_{j=1}^5 z_j^{M_{ji}}
\end{equation} 
of the degree $\bar{d}=\sum_j \bar{k}_j M_{ji}$ defines the Calabi-Yau hypersurface  $X_{M^T}$ in  another projective space 
$\mathbb{P}^4_{(\bar{k}_1,\bar{k}_2,\bar{k}_3,\bar{k}_4,\bar{k}_5)}$.
The weights $\bar{k}_i$ satisfy the Calabi–Yau condition
 $\sum_i \bar{k_i}=\bar{d}$. 
 
Taking the quotient $G^T:=G_0^T/J_{M^T}$, we define the Calabi-Yau orbifold as
\begin{equation}
    Z(M^T,G^T):=X_{M^T}/G^T.
\end{equation}
The fact is that groups $G$ and $G^T$ can be chosen in different ways.
The question arises: is it possible to choose a group $G^T$ for a given group $G$, and if so,  how to do this so that the manifolds $Z(M,G)$ and $Z(M^T,G^T)$ form a mirror pair? 

The answer was given by Krawitz \cite{krawitz2009fjrw}. His construction allows one to define the generators of the group $G_0^T$. Namely, they are constructed using the exponents $S_{li}$ of the invariant monomials as (\ref{eq:Monomials}) as follows
\begin{equation}
    \rho_l^T:=\prod_{i=1}^5 q_i(M^T)^{S_{li}},
\end{equation}
where $q_i(M^T)$ are generators of the $\operatorname{Aut}(M^T)$ acting on each coordinate $z_j $  in $\mathbb{P}^4_{(\bar{k}_1,\bar{k}_2,\bar{k}_3,\bar{k}_4,\bar{k}_5)}$ as
\begin{equation}
    q_i(M^T): z_j \mapsto e^{2\pi i B_{ij}}z_j.
\end{equation}
It follows that the group $G_0^T$ acts on the coordinates $z_j $ as
\begin{equation}
\label{eq:HTaction}
    \rho_l^T: z_j \mapsto e^{2 \pi i \sum_i S_{li}B_{ij}} z_j.
\end{equation}
The full family of the mirror Calabi-Yau orbifold $Z(M^T,G^T)$ is given by zero locus of 
\begin{equation}
\label{eq:CalabiYauMirror}
    W^{Z(M^T,G^T)}(z,\psi)= \sum_{i=1}^5 \prod_{j=1}^5  z_j^{M_{ji}}+ \sum_{m=1}^{h_Y} \psi_m \prod_{j=1}^5 z_j^{R_{mj}},
\end{equation}
where $\psi_m$ are moduli of complex structure of the family $Z(M^T,G^T)$. The monomials $\bar{e}_m:=\prod_{j=1}^5 z_j^{R_{mj}}$ are invariant under the $G_0^T$ action (\ref{eq:HTaction}):
\begin{equation}
\label{eq:invcond}
    \rho_l^T \cdot \bar{e}_m = e^{2 \pi i \sum_{ij} B_{ij}S_{li}R_{mj}} \prod_{j=1}^5 z_j^{R_{mj}} = \prod_{j=1}^5 z_j^{R_{mj}}.
\end{equation}
The exponents $S_{li}$, and $R_{mj}$ can be interpreted as a five-component integer vectors $(\vec{S}_l)_j=S_{lj}$, and $(\vec{R}_m)_i=R_{mi}$. 
The invariance condition (\ref{eq:invcond})  can be rewritten in terms of the pairing of these vectors, defined with matrix $B$ as
\begin{equation}
\label{eq:pairing}
    (\vec{S}_l,\vec{R}_m)=\sum_{i,j=1}^5 B_{ij}S_{li}R_{mj} \in \mathbb{Z}.
\end{equation}
This relation is a strong restriction because although the matrices $S$ and $R$ are integers, at the same time the entries in $B$ are rational.
Taking into account also the condition of quasi-homogeneity 
\begin{equation}
    \label{eq:homogmirror}
    \sum_{i=1}^5R_{mi}\bar{k}_i=\bar{d},
\end{equation}
we conclude that the equations (\ref{eq:pairing}) have finite non-negative number of solutions. Denote this   number as $h_Y$.

The Chiodo-Ruan theorem  \cite{aless2009lgcy} states that orbifolds $Z(M,G)$ and 
$Z(M^T,G^T)$ form a mirror pair on the level of cohomology
\begin{equation}
  H^{p,q}(Z(M,G),\mathbb{C})=H^{3-p,q}(Z(M^T,G^T),\mathbb{C}).
\end{equation}
Thus the Berglund–Hübsch–Krawitz construction allows one to determine 
the polynomial $W^{Z(M^T,G^T)}$ which defines the full family of $Z(M^T, G^T)$. 

\section{Batyrev construction and verification of equivalence}
\label{sec:3}
In this section, following the Batyrev's approach 
\cite{Batyrev:1994hm}, briefly described in the section \ref{sec:intro}, 
we will  build the mirror $Y$ of  the orbifold  $Z(M,G)$ as 
 a hypersurface in toric variety. 
 We find the mirror polynomial $W^Y$ as a function of toric coordinates. We also show that the mirror hypersurface $Y$  is equivalent to the Calabi-Yau orbifold $Z(M^T,G^T)$ (\ref{eq:CalabiYauMirror}) in a weighted projective space obtained above using the BHK construction. We do this in two steps.

The first step is to construct the toric variety. To do this, we begin with the invertible quasi-homogeneous polynomial $W_M$ 
(\ref{eq:Familly}) with the deformations $e_l$. 
Rewrite it  in a form
\begin{equation}
    W^X(x,\varphi)=\sum_{i=1}^5 \prod_{j=1}^5 x_j^{M_{ij}}+\sum_{l=1}^{h_X}\varphi_l \prod_{j=1}^5 x_j^{S_{lj}}= \sum_{a=1}^{h_X+5}C_a \prod_{j=1}^5 x_j^{V_{aj}}.
\end{equation}
The exponents $V_{aj}$ are components of integer vectors, namely $V_{aj}=(\vec{V}_a)_j$. They are equal:
\begin{equation}
V_{a j}=\left\{\begin{array}{ll}
M_{a j}, & 1 \leq a \leq 5 \\
S_{a-5, j}, & 6 \leq a \leq h
\end{array}\right.
\end{equation}
and $\Vec{V}_{h_X+5}$ has components $(1,1,1,1,1)$.
  From the condition 
\begin{equation}
\label{eq:condV}
\sum_{i=1}^5V_{ai}k_i=d
\end{equation}
it follows, that the vectors 
$\Vec{V'}_a :=\Vec{V}_a - \Vec{V}_{h_X+5}$ lie in the four-dimensional lattice  $L \subset \mathbb{Z}^5$. More precisely, they belong to the sublattice of $L$ defined by the group $G$. 

These vectors $\vec{V'}_a$  correspond to the lattice points of the reflexive polytope $\Delta(M,G)$. They also correspond to the edges of the fan of the reflexive polyhedron dual to the original one \cite{Batyrev:1994hm,Hori:2003ic}.  
The fan defines the toric variety $T$. Let us construct this toric variety. The vectors $\vec{V'}_a, \ a=1,\dots, h_X+4$, being four-dimensional,   satisfy to $h_X$   linear relations
\begin{equation}
\label{eq:toricrelations}
    \sum_{a=1}^{h_X+4}Q_{la}\vec{V'}_a=0, \ \ \ l=1,\dots,h_X,
\end{equation}
where the coefficients $Q_{la}$ are integers.
The solution of the equations (\ref{eq:toricrelations}) is
\cite{Aleshkin:2018tbx}
\begin{equation}
\label{eq:Q}
Q_{l a}=\left\{\begin{array}{ll}
S_{l j} B_{j a}, \ & 1 \leq a \leq 5 \\
-\delta_{l, a-5}, \ & a>5.
\end{array}\right.
\end{equation}
The numbers $Q_{l a}$ are the weights of  this toric variety 
$T$. Knowing  them allows one to define $T$ as \cite {Hori:2003ic} 
\begin{equation}
    \label{eq:toricvar}
    T=\frac{\mathbb{C}^{h_X+4}-Z}{(\mathbb{C}^*)^{h_X}},
\end{equation}
where $Z$ is the $(\mathbb{C}^*)^{h_X}$--invariant subset 
\cite {Hori:2003ic}. The abelian group $(\mathbb{C}^*)^{h_X}$ acts on the coordinates in $T$ as
\begin{equation}
    y_a \mapsto \lambda^{Q_{la}} y_a, \ \ \ l=1,\dots,h_X;\  a=1,\dots,h_X+4.
\end{equation}

The mirror Calabi-Yau manifold $Y$ is realized as a hypersurface in the toric variety $T$, cut out by an invariant under the action of the abelian group $(\mathbb{C}^*)^{h_X}$ 
homogeneous polynomial $W^Y(y)$ \cite{Aleshkin:2018tbx, Belavin:2019gbx}
\begin{equation}
\label{eq:conditionY}
    W^Y(\lambda^{Q_{l1}}y_1,\dots,\lambda^{Q_{l h_X+5}}y_{h_X+4})=W^Y(y_1,\dots,y_{h_X+4}), \ \ \ l=1,\dots,h_X-1,
\end{equation}
\begin{equation}
\label{eq:newconditionY}
    W^Y(\lambda^{Q_{h_X,1}}y_1,\dots,\lambda^{Q_{h_X, h_X+5}}y_{h_X+4})=\lambda W^Y(y_1,\dots,y_{h_X+4}), 
\end{equation}
We are looking for  the $ W^Y$ as a sum quasi-homogeneous monomials
\begin{equation}
\label{eq:toricpolynon}
    W^Y(y)=\sum_{b=1}^{h_Y+5}\tilde{C}_b \prod_{a=1}^{h_X+4}y_{a}^{N_{ba}}.
\end{equation}
The exponents $N_{ba}$ are non-negative integer numbers. 
We know 5 invariant monomials of the form
\begin{equation}
\label{eq:monomialsintoric1}
        P_i  =\prod_{a=1}^{h_X+4}y_a^{V_{ai}}=\prod_{j=1}^{5}y_j^{M_{ji}}\times \prod_{l=1}^{h_X-1}y_{l+5}^{S_{li}}=\prod_{j=1}^{5}y_j^{(M^T)_{ij}}\times \prod_{l=1}^{h_X-1}y_{l+5}^{S_{li}},
\end{equation}  
where $i=1,...,5$.  
We can check that  monomials $  P_i $  are invariant under
 the $(\mathbb{C}^*)^{h_X}$ action, because $\sum_a Q_{la}V_{ai}=0$. 
Also, we have one more invariant monomial of the form
\begin{equation}
    \bar{e}_{h_Y}=\prod_{a=1}^{h_X+4}y_a.
\end{equation}
Other invariant monomials we  seek in the form
\begin{equation}
\label{eq:monomialsintoric2}
       \bar{e}_m  =\prod_{a=1}^{h_X+4}y_a^{N_{ma}}=\prod_{j=1}^5y_j^{N_{mj}}\times \prod_{l=1}^{h_X-1} y_{l+5}^{N_{m,l+5}}.
\end{equation} 

From the invariance of the  $\bar{e}_m$ 
 under $(\mathbb{C}^*)^{h_X}$ action one can obtain that the non-negative integers $N_{mi}$ need to satisfy the relations
\begin{equation}
\label{eq:eqint}
   B_{ji}S_{lj}N_{mi}-N_{m,l+5} =\delta_{l,h_X}.
\end{equation}
Since the numbers in the r.h.s. $N_{m,l+5}$ are assumed to be integers,  we get that the  integer numbers $N_{mi}$ satisfy  the relation
\begin{equation}
\label{eq:NinZ}
B_{ji}S_{lj}N_{mi} \in \mathbb{Z},
\end{equation}
which exactly coincides with the relation for the exponents $R_{mi}$ in (\ref{eq:pairing}).\\
Using the fact that  $S_{h_X,j}=1$  and the relation 
(\ref{eq:matrixBT}) for the matrix $B$ we get from   (\ref{eq:eqint}) 
for $l=h_X$ 
\begin{equation}
\label{eq:condN}
    N_{mj}\bar{k}_j=\bar{d}.
\end{equation}
Therefore, as it  follows  from the (\ref{eq:NinZ}) and (\ref{eq:condN})  that these equations  have the same  solutions as the equations
 (\ref{eq:pairing}) in the BHK approach, so that 
\begin{equation}
    N_{mi}=R_{mi}.
\end{equation}
Now we can find the remaining exponents. Using the (\ref{eq:eqint}) we obtain that  
\begin{equation}
\label{eq:m,l+5}
    N_{m,l+5}=S_{lj}B_{ji}R_{mi}.
\end{equation}
 Thus, we finally obtain that the polynomial $W^Y$ has the form
\begin{equation}
\label{eq:WYtoric}
     W^Y(y_1,\dots,y_{h_X+4})=\sum_{i=1}^5\prod_{j=1}^{5}y_j^{M_{ji}}\prod_{l=1}^{h_X-1}y_{l+5}^{S_{li}}+\sum_{m=1}^{h_Y} \psi_m \prod_{j=1}^{5}y_j^{R_{mj}}\prod_{l=1}^{h_X-1}y_{l+5}^{(S_{lj}B_{ji}R_{mi})}.
\end{equation}

Then, the mirror Calabi-Yau manifold $Y$ is given by zero locus of $W^Y$ (\ref{eq:WYtoric}) in $T$, similar as it was in  \cite{Witten:1993yc}. 
To see the correspondence of the just described construction to the previous BHK construction, we introduce the new coordinates 
\begin{equation}
\label{eq:map}
    (y_1,\dots,y_{h_X+4})\mapsto (z_1,z_2,z_3,z_4,z_5),
\end{equation}
where
\begin{equation}
\label{eq:projcoord}
    z_j=y_j\prod_{l=1}^{h_X-1}y_{l+5}^{S_{li}B_{ij}}, \ \ \  1 \leqslant z_j \leqslant 5.
\end{equation}
The change of the coordinates (\ref{eq:map}) is mapping of the toric variety $T$ into the quotient 
of the weighted projective space $\mathbb{P}^4_{(\bar{k}_1,\bar{k}_2,\bar{k}_3,\bar{k}_4,\bar{k}_5)}/G^T$.
Indeed, the abelian group 
$(\mathbb{C}^*)^{h_X}$ acts on $z_j$ as
\begin{equation}
    z_j \mapsto \lambda^{Q_{sj}+Q_{s,l+5}S_{li}B_{ij}} z_j=\lambda^{S_{sm}B_{mj}-\delta_{sl}S_{li}B_{ij}}z_j=z_j, \ \ \ s=1,\dots,h_X-1.
\end{equation}
And, in view of the fact  that $S_{h_Xj}=1$ and the property of the matrix $B$ (\ref{eq:matrixBT}),
\begin{equation}
    z_j \mapsto \lambda^{S_{h_X m}B_{mj}} z_j = \lambda^{\sum_m B_{mj}}z_j=\lambda^{\nu_j}z_j,
\end{equation}
 where $\nu_j:=\bar{k}_j/\bar{d}$.

It is easy to check, that the polynomial $W^Y$ can be rewritten in these coordinates as
\begin{equation}
     W^Y(z)=\sum_{i=1}^5\prod_{j=1}^{5}z_j^{(M^T)_{ij}}+\sum_{m=1}^{h_Y} \psi_m \prod_{j=1}^{5}z_j^{R_{mj}}
\end{equation}
and coincides with the result of Berglund–Hübsch–Krawitz construction (\ref{eq:CalabiYauMirror}).

Thus, in this paper, we have provided a simple evidence of equivalence of two mirror partners obtained by Berglund–Hübsch–Krawitz, and Batyrev constructions. Namely started with an invertible polynomial $W_M$, suitable subgroup $G$ and the weighted projective space $\mathbb{P}^4_{(k_1,k_2,k_3,k_4,k_5)}$ we have obtained the mirror in another projective space $\mathbb{P}^4_{(\bar{k}_1,\bar{k}_2,\bar{k}_3,\bar{k}_4,\bar{k}_5)}$. Following to Batyrev approach, we have constructed the toric variety and defined the mirror polynomial $W_M$. Finally, we showed the connection between the toric variety and the weighted projective space. The results of these two mirror constructions coincide.
Earlier, we checked a similar statement for the particular case of quintic orbifolds in \cite{BelavinEremin2}.

\section*{Acknowledgments} 

We are grateful to V. Belavin, M. Bershtein and G. Koshevoy for useful discussions. This work was performed in Landau Institute for Theoretical Physics. The work of A. Belavin was done within the framework of the state assignment. The work of B. Eremin has been supported by the Russian Science Foundation under the grant 18-12-00439.


\end{document}